# The Golden Gate Textile Barrier: Preserving California's Bay of San Francisco from a Rising North Pacific Ocean

Richard B. Cathcart
Geographos
Burbank, California
USA
E-mail: rbcathcart@charter.net

Alexander A. Bolonkin
C & R
1310 Avenue R
Suite 6-F
Brooklyn, New York 11229
USA
E-mail: abolonkin@juno.com

**Abstract**

Climate change in California may require construction of a barrier separating the Pacific Ocean from San Francisco Bay and the Sacramento River-San Joaquin River Delta simply because Southern California is remarkably dependent on freshwater exported from the Delta. We offer a new kind of salt barrier, a macroproject built of impermeable textile materials stretched across the Golden Gate beneath the famous bridge. We anticipate it might eventually substitute for a recently proposed "San Francisco In-Stream Tidal Power Plant" harnessing a 1.7 m tide at the Bay's entrance if future climate conditions Statewide is conducive. First-glance physics underpin our macroproject.

**1. Introduction**

While the public in Europe and the USA apparently have adopted a common outlook of technical resignation towards the world's evident post-Ice Age rising ocean, macroengineers in Dubai assiduously sculpt the seashore and create new inhabitable islands (Koolhaas, 2006; Marreiros, 2006). Indeed, by emplacement of a Strait of Hormuz Dam, Macro-engineering's foremost proponents have formulated a workable plan to isolate the Persian Gulf from the ocean (Schuiling et al., 2005) and, thereby, making the Persian Gulf's water level controllable. It is true that macroprojects protecting major European cities from storm-surges—such as London (Lavery and Donovan, 2005), St. Petersburg (Klevanny and Smirnova, 2002) and Venice (Fletcher and Spencer, 2005)—have been constructed or continue to be considered. The in-place infrastructures of The Netherlands are well known (Smits et al., 2006)! In Asia, a movable textile seawater barrier facility is proposed to shield Palk Bay, located between India and Sri Lanka, from infrequent tsunami and yearly storm surges (Cathcart, 2006). To "correct oceanographic problems impairing the economic usefulness of coastal land", Richard B. Cathcart and Alexander A. Bolonkin (2007) offered a deliberate "terracing" of the Mediterranean Sea within its vast Basin by its enclosure with a textile barrier at the Strait of Gibraltar that controllably excludes Mediterranean Sea evaporation compensation Atlantic Ocean seawater inflow. Such comprehensive and geographically large-scale corrective macroproject concepts stem, in part, from stimulation initiated by a summarizing technical book about Macro-engineering's burgeoning professionalism (Badescu et al., 2006).

## 2. Good cause for concern by coastal Americans

Before Macro-engineering and macro-management can be considered in the context of California's coast, a fundamental question must be asked: Is sea level rising along the USA's coast? After 1970, the world's news media discovered "global warming" and, since then, many prominent Californian legislators and politicians, such as Governor Arnold Schwarzenegger during 2007, as well as other citizens are nowadays convinced that anthropogenic carbon dioxide and other greenhouse gas global emissions are directly responsible for raising the Pacific Ocean's level of impingement on the State's coastline. Still, the tandem link between enhanced global greenhouse gas emissions and global sea level rise is still an unresolved matter of geoscientific dispute (Larsen and Clark, 2006). As Earth's most recent Ice Age wanes, the world's atmosphere is likely to warm, some glaciers will become smaller and the ocean impacting the USA's coast will probably rise to an elevation noticeably above the present-day's level.

## 3. East Coast: Metropolitan New York City sea level rise impacts

Storm-induced seawater surges, caused by hurricanes and nor'easters, have in the past impacted the metropolitan New York City region. Future local sea level rise is expected to boost the destructive power of such storms to disrupt the function of vital urban infrastructure and to injure and kill people (Gornitz, 2002). Consequently, the possibility of protecting the Metropolitan Region with storm-surge barriers is being studied. Closable steel and concrete storm-surge barriers erected in three straits (between New York Bay and Lower New York Bay, Arthur Kill and near the Throngs Neck Bridge) could effectively, though only temporarily, seal off the Metropolitan Region from incoming Atlantic Ocean-generated storm-surges (Bowman et al., 2005).

## 4. Gulf Coast: post-2005 New Orleans Region recovery

By the end of 2007, the US Army Corps of Engineers must present to Congress a long-term, comprehensive plan for hydraulic macroprojects that will offer Gulf Coast residents and industry a predictably successful future defense against hurricane storm-surge damage (Sparks, 2006). A complex integrated system of walls, barriers, offshore breakwaters, barrier island beach dune restoration, repaired or new levees, and massive mobile mechanical barrages that would close across inlets to exclude surging hurricane-stirred seawater and freshwater is being considered. Americans have modified the Gulf Region's coast, either directly, by construction or dredging, or indirectly, as a consequence of profound inland landscape changes that then influence future sediment supply, freshwater runoff, or the Gulf Region's climate (Shallat, 2006).

The commonest settlement/use practices that significantly alter the USA's coast are the construction of coast-protection infrastructures such as jetties, groins, seawalls, bulkheads, revetments and the development of private and public property on and close to the shoreline. These oceanographic infrastructures constitute "Hard Structure Armoring". The alternative is "Beach Nourishment". Beach nourishment is the only coastal macro-management technique that adds sand to the littoral system nearest the seashore. Generally speaking, "Hard Structure Armoring" is monetarily the costliest—about 10% of initial installation cost annually—to maintain in a robust state of physical readiness.

## 5. West Coast: San Francisco Bay/Delta Region

Last year, the U.S. Geological Survey published a useful geographical and historical survey of California's coastal land gains and losses (Hapke et al, 2006). The present-day post-Ice Age sea level rise at the entrance to the 890 square kilometer Bay of San Francisco is ~2.29 cm/decade. Future sea level rise, and the subsequent elevated storm-surges thereon, may severely affect San Francisco Bay shoreline cities, the levee-enclosed low-lying land above the intertidal flats (Jaffee and Foxgrover, 2006) and, very importantly, the adjacent Sacramento River/San Joaquin River Delta.

Massive seawater intrusions into the largest bay on California's coast, its wetlands, associated surface freshwater storage systems and groundwater aquifers are undesirable prospective degradations of the San Francisco Bay/Delta. For example, it is imaginable that the Harvey O. Banks Pumping station feeding the California Aqueduct and the Tracy Pumping Plant feeding the Delta-Mendota Canal would have to cease operation, at least temporarily. Contamination of the Delta's freshwater with a permanent influx of saltwater would endanger a massive public freshwater supply that is pumped to Southern California!

"Projected sea level rises of 20-80 cm…during the twenty-first century can only be expected to compound the vulnerability of subsided Delta Islands to levee failure…and increase upstream backwater flooding" (Cayan et al., 2006). The Delta receives runoff from about 40% of California's land area (163,000 square kilometers) and about 50% of the State's total streamflow. A great earthquake at the infamous nearby San Andreas Fault—more powerful than the 7.1 Richter scale "Loma Prieta" earthquake of 17 October 1989—could cause many kilometers of Delta levees to collapse. Widespread Delta levee failure would divert river flow into many Delta polders—perhaps as much at 2.5 to 3.0 billion cubic meters (Mount and Twiss, 2005)—and, consequently, generate a strong flow of seawater from San Francisco Bay eastward into the unprotected Delta, suddenly contaminating the freshwater normally pumped to Southern California's populace! The volume of San Francisco Bay is ~6.165 billion cubic meters. It has a mean depth of ~6 m.

In addition, a nearly inevitable increase of human population on San Francisco Bay's periphery is expected to require desalination plants providing a reliable future supplemental freshwater supply (Yeung, 2005). Social fabrics and vital infrastructure will be stressed and strained negatively (Gleick and Maurer, 1980; Hayhoe et al., 2004). Furthermore, future industrial accidents may cause the generation of an infrastructure-devastating tsunami that might reach inland seaports such as Sacramento and Stockton, especially at high tide or during periods of storm-surge (Greenberg, 1993). Though computer models predict the $21^{st}$ Century's climate for California will be dryer than today, it is worth noting that, during January 1862, the "…outflow of fresh water into the bays of San Francisco and into the Pacific Ocean through the Golden Gate was both large and persistent. Sea level at the Golden Gate was 17 cm above normal…. For nearly two weeks fresh water flowed continually seaward through the Golden Gate, without tidal fluctuation. Fresh water covered the surface of the…[sub-embayments] for two to three months" (Engstrom, 1996). Peterson et al (1985) estimated that a freshwater inflow of ~120,000 $m^3$/second would have been necessary to overcome the tides [tidal range: maximum 2.65 m, ~2.0 billion cubic meters moving at ~2.5 m/second every 6 hours] at the Golden Gate during January 1862! In other words, an identical flood flow rate today could fill San Francisco Bay entirely with freshwater in only 14.27 hours!

What if California's climate becomes wetter than today? In that eventuality, more freshwater may be exported to Southern California. But if, as predicted, California becomes drier, then San Francisco Bay may need to be dammed at the Golden Gate to forestall saltwater intrusion into the Delta. The entrance to San Francisco Bay, the deepest in the bay (107 m), is an elongate seafloor depression with steep sidewalls of $10^0$ to $17^0$. The ocean shoals immediately west of the Golden Gate to a depth of ~38 m. FIGURE 1, below, is a computer simulation of the seafloor in the vicinity of the Golden Gate Bridge. Farther out to sea, a huge arcuate sandbar is present which is a relict from the Ice Age. Immediately after the 18 April 1906 earthquake, a tidal anomaly at the Presidio's tide-gauge station was registered that indicated a 10 cm lowering of the sea level for a quarter of an hour followed by two to three tidal oscillations with maximum amplitude of only 5 cm (Geist and Zoback, 2002).

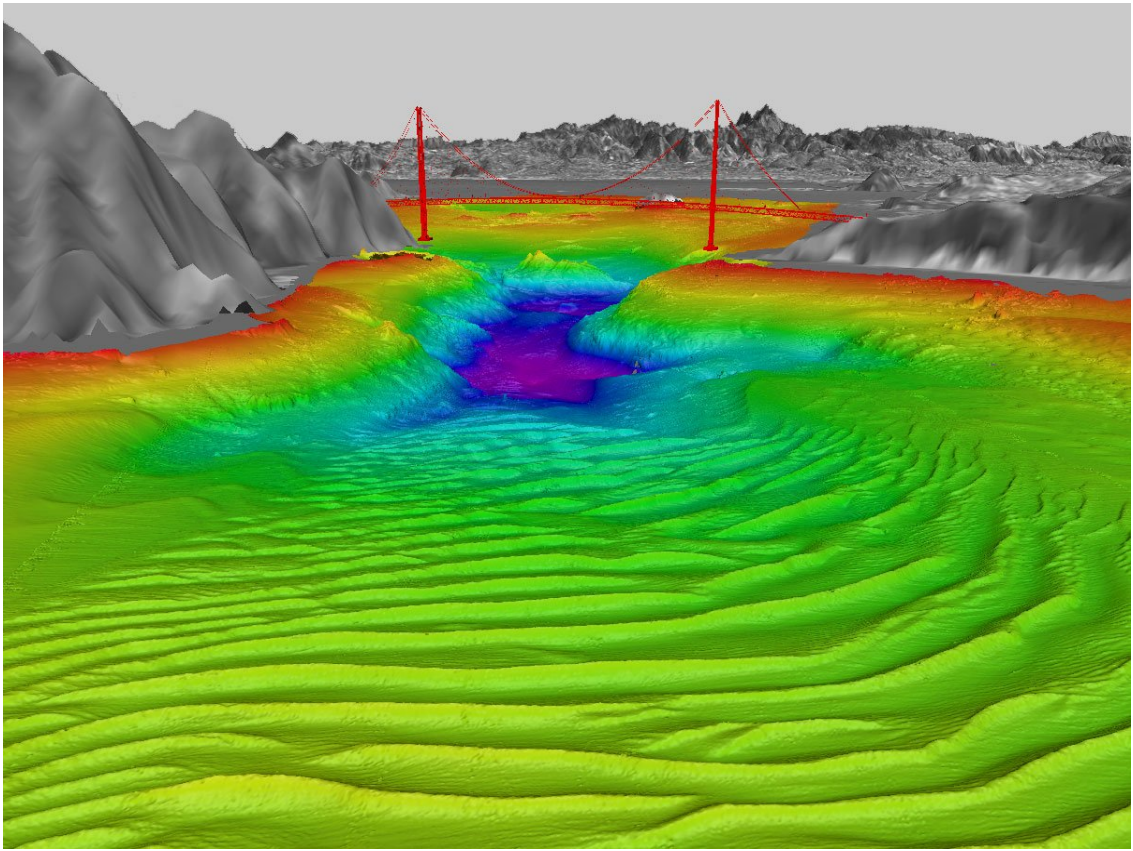

**Fig. 1.**

**6. The Region's planned, but unbuilt, Hard Structure Armoring**

Rivers first passed seaward through the Golden Gate approximately 500,000 years ago and post-Ice Age sea level first rose high enough to flow into the Bay of San Francisco about 15,000 years ago (Dartnell et al., 2006). The northern reach is a partially mixed sub-embayment dominated by seasonally varying freshwater inflows while the south reach is a tidally oscillating estuary. The Bay of San Francisco is an ocean-river mixing zone with a seaward flow equal to the sum of the river inflows less precipitation; at the present time only the freshwater inflow can be subjected to macro-management. During droughts, freshwater inflows have been, effectively, nil; droughts have produced saltwater intrusions of the Sacramento River-San Joaquin River Delta. Anti-saltwater intrusion barriers were first proposed by nascent macroengineers during the 1920s as a means of halting the contamination of Delta farmland irrigation freshwater as well as that used by nearby industries.

Since about 1950—the year when the term "coastal engineering" was first used in print in the **Proceedings of the First Conference on Coastal Engineering**—most of San Francisco Bay has shifted from being depositional to erosional as sediment supply diminished (when dams were built inland) and existing currents and waves continued to remove sediment from the Region. Essentially all of San Francisco Bay has an anthropogenic character (Chin et al, 2004). Any future rise of the Pacific Ocean will, therefore, impact all existing and planned shoreline infrastructures. In 1990, landscape architect Steven Garey Abrahams proposed "Landscape Mobilis" which would consist of a fleet of waterborne landscapes—gardens, forests, meadows, crops, vineyards on barges towed around San Francisco Bay, visiting "receptacle parks" at strategic places along the shore (Raver, 1991).

When saltwater intrusions of the Delta first became a remarkable macro-problem for commercial farming and industry in the Delta during the early 1900s, macroengineers of the time proposed physical barriers to block the inland migration of salty water. Many salt barrier plans were formulated, most of which involved the construction of low-level dams to separate freshwater upstream from the tidal seawater on the downstream side. Following the very low Delta outflow of 1923-1924, a 1929 macroproject plan for a barrier built at the Golden Gate was the greatest in geographical scope ever promoted by reclamation and hydraulic experts. Offered by Walker R. Young, and had it been constructed, the dam would have become the longest "Hard Structure Armoring" of California's coast ever conceived and, so far, discussed (Young, 1929)! FIGURE 2, below illustrates Young's concept.

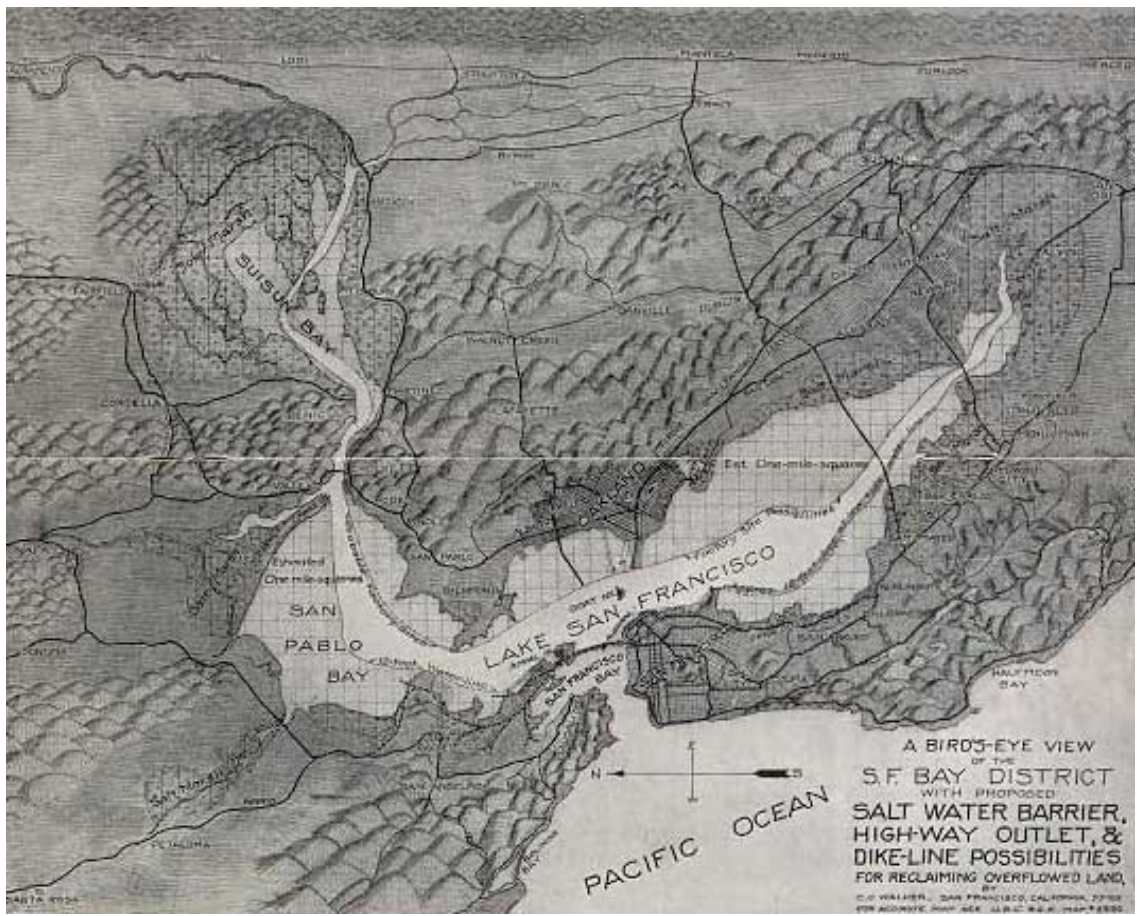

**Fig. 2**.

It would have been unnecessary to fabricate the Golden Gate Bridge, completed in 1938, if W.R. Young's 1927-29 enclosure work and super-highway had ever been constructed!

Walker Rollo Young (1885-1982), a 1908 University of Idaho graduate of engineering, became Chief Engineer at the USA's Bureau of Reclamation in 1945 after serving, from 1930, as Construction Engineer at Hoover Dam. Due to its site, W.R. Young's "Golden Gate Dam and Highway" would have made the small embayment west of his proposed saltwater barrier a collection place for material being moved by wave and current action in the littoral cells to the north and south of the entrance to an areally smaller, enclosed Bay of San Francisco (Barnard, 2005; d'Angremond and van Roode, 2004). Maintenance dredging would have become necessary. Basically, his "Golden Gate Dam and Highway" would have shortened the fetch of winds driving waves into San Francisco Bay—such winds are strongest during Summer and during Winter storms—and since San Francisco Bay was to be a freshwater lake—one might say, a "real-estate lake" (Rickert, 1971)—the ecology of the Region would duplicate that endured by the Region during January 1862!

Subsequent to Young's concept, from 1933 until is death, John Reber (1899-1960) campaigned for his plan to convert ~85% of the bay into a freshwater lake by closing off the bay's northernmost and southernmost reaches and adding housing and industrial sites to the artificial shore. Similar macroprojects, with different geographical and economical goals, were offered by Norman Sper, who suggested that the Hudson River between New York Harbor and the Harlem River be filled with rock and soil, thereby making Manhattan Island no longer an island (Albelli, 1934). Imagine "Tetrahedral City" invented by R. Buckminster Fuller (1895-1983) floating peacefully on Walker R. Young's vision of Lake San Francisco (Fuller and Marks, 1973)!

**7. Opportunity for future Golden Gate Textile Barrier (GGTB)**

The GGTB is worthy of investigation at this time because the City and County of San Francisco is considering the installation of a tidal in-stream hydroelectric permeable "curtain" spanning the Golden Gate near the present-day bridge connecting San Francisco and Marin Counties of California. Gulf Stream Energy, Inc. and Golden Gate Energy Company's Joint Application for Preliminary Permit for the San Francisco Bay Tidal Energy Project under P-12585-000 was filed on 26 April 2005 with the Federal Energy Regulatory Commission in Washington, D.C.

On 10 June 2006 a thorough technical examination of such a system for submarine power generation—that is, its system design, potential performance, construction and operational costs, as well as its economic and environmental benefits by the Electric Power Research Institute Inc. (EPRI) was publicly displayed (Previsic, 2006). EPRI's experts concluded that ~35 MW could be extracted from the tidal stream "…without any negative impact on the environment" at a facility running North-South along the $122.47836^0$ West meridian from $37.8111^0$ North (on the San Francisco side of the strait) to $37.8265^0$ North (on the Marin County side of the strait), a distance of ~1,380 m where the mean depth is ~54 m. Since the City and County of San Francisco has an estimated average electrical power consumption of ~570 MW it is, therefore, possible that approximately 6% of the City and County's electrical supply could be developed by an underwater tidal energy "permeable curtain". Obviously, this nearly $100 million facility will not dam the Golden Gate nor will it impede the navigation of vessels entering or leaving the Bay of San Francisco, California. What it seems to indicate is (1) some influential people and organizations in San Francisco will tolerate an invisible hydropower plant at the Golden Gate and (2) future macroengineering improvements there might also be tolerated if the Pacific Ocean should rise and/or the runoff passing through the Sacramento River-San Joaquin River Delta is greatly reduced by State-wide future climate change. The great seafloor pit below the Golden Gate Bridge may prove to be a boon since it can hold a great deal of sediment that may shift westwards from the central part of the Bay of San Francisco. And, if a Golden Gate Textile Barrier is emplaced, sand

naturally migrating in the shallow Pacific Ocean's nearest littoral cell may improve the tsunami run-up on the facing cliffed-coastline that is present north and south of the entrance to San Francisco Bay. The GGTB would likely be equipped with penetrating ship locks, perhaps emulating the design offered by Martin Cullen in his U.S. Patent Application 20050163570 issued 28 July 2005.

**8. Seawater/freshwater textile tension structure**

The typical textile barrier—more accurately, a membrane—is shown in FIGURE 3, below. FIGURE 3 includes the textile membrane seawater/freshwater barrier, floats, underwater pump/electricity generation, and the support cable. The hydropower dam can safely operate even when the water level on one side is as much at 10 m.

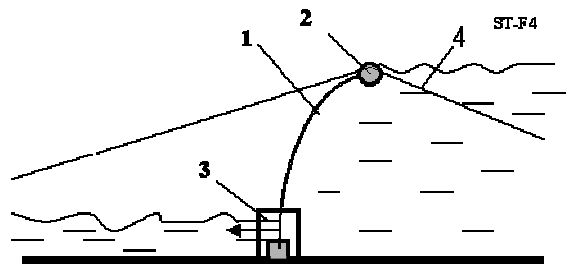

**Fig. 3.** Textile Barrier. Notations: 1 - textile membrane, 2 - floats, 3 - pump/electric station, 4 - support cable.

# 8.1. Theory and Computation

1. **Relative concentration of salt in water.**

When a river's freshwater inflows to a closed estuary that passes the freshwater only in one direction (to ocean), the salt concentration of the seawater decreases. New relative concentration may be computed by equation

$$\overline{c} = e^{-Qt} , \qquad (1)$$

where $Q$ is a relative inflow of a river freshwater. $Q = q/W$ where $q$ is inflow of freshwater, m³; $W$ is volume of like (golf), m³; $t$ is time, sec. Computation for this equation is illustrated below

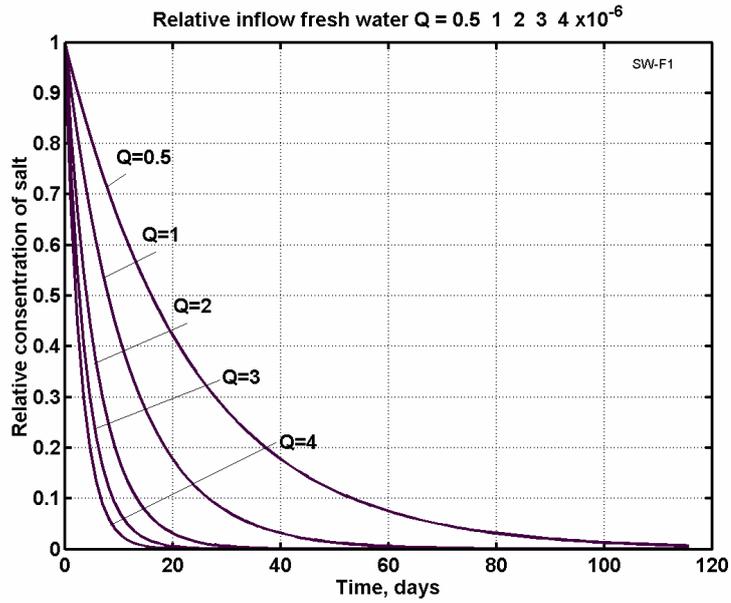

**Fig. 4.** Relative concentration of salt via time (days) for different relative inflow of a river freshwater $Q = q/W$ where $q$ is inflow of fresh water, m³; $W$ is volume of estuary, m³.

2. **Force** $P$ [N/m²] for 1 m² of dam is
$$P = g\gamma h, \qquad (2)$$
where $g = 9.81$ m/s2 is the Earth's gravity; $\gamma$ is water density, $\gamma = 1000$ kg/m3; $h$ is difference between top and lower levels of water surfaces, m (see computation in FIGURE 5).

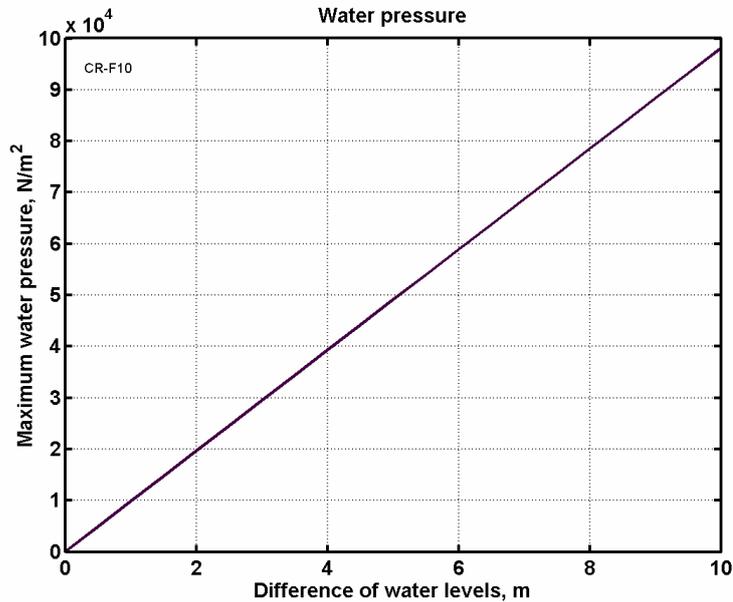

**Fig. 5**. Water pressure via difference of water levels

3. **Water power** $N$ [W] is

$$N = \eta g m h, \quad m = \gamma v S, \quad v = \sqrt{2gh}, \quad N = \eta g \gamma h S \sqrt{2gh}, \quad N/S \approx 43.453\eta h^{1.5}, \quad [\text{kW/m}^2] \qquad (3)$$

where *m* is mass flow across 1 m width kg/m; *v* is water speed, m/s; *S* is turbine area, m²; *η* is coefficient efficiency of the water turbine, *N/S* is specific power of water turbine, kW/m².
Computation is presented in FIGURE. 6.

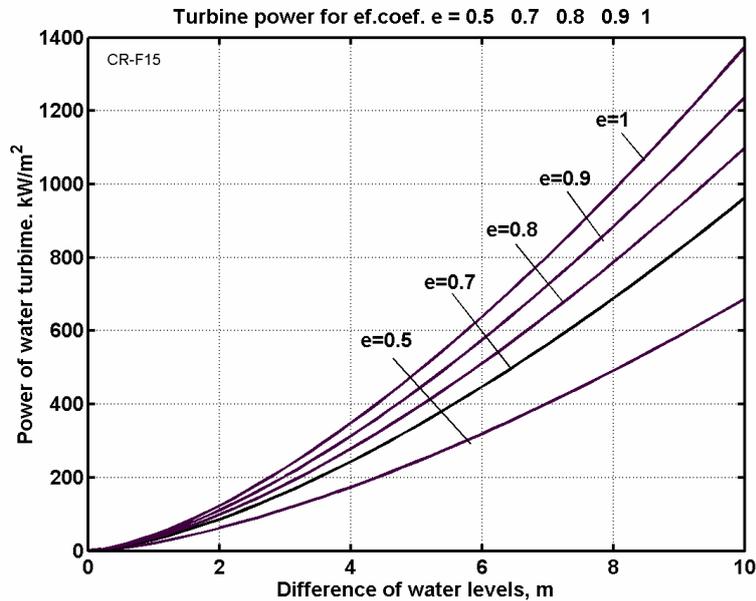

**Fig. 6**. Specific power of a water turbine via difference of water levels and turbine efficiency coefficient

**4. Film thickness** is

$$\delta = \frac{g\gamma h^2}{2\sigma},\qquad(4)$$

where σ is safety film tensile stress, N/m². Results of computation are in Fig. 7. The fibrous material (Fiber B, PRD-49) has maximum $\sigma = 312$ kg/mm² and specific gravity $\gamma = 1.5$ g/cm³.

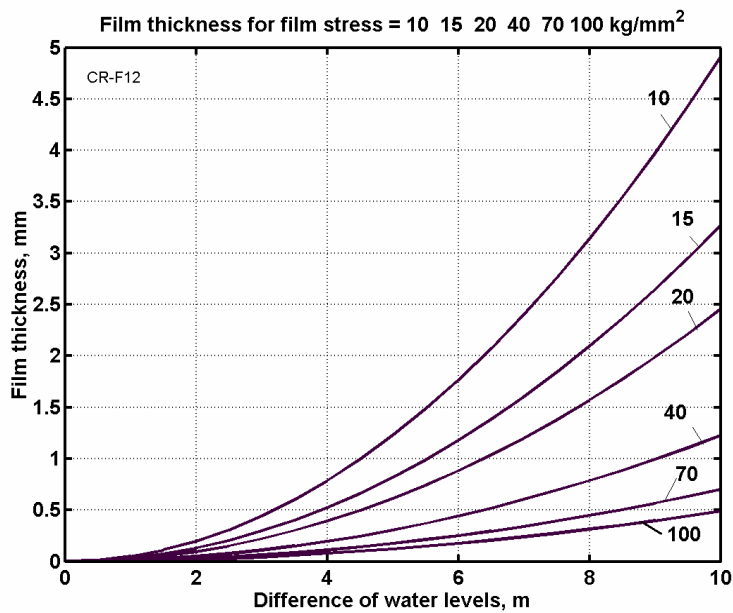

**Fig. 7**. Film (textile) thickness via difference of water levels safety film (textile) tensile stress.

**5. The film weight** of 1m width is

$$W_f = 1.2\delta\gamma H, \qquad (5)$$

Computation is illustrated by FIGURE 8. If our proposed textile barrier has a length $L$ m, we must multiple this result by $L$.

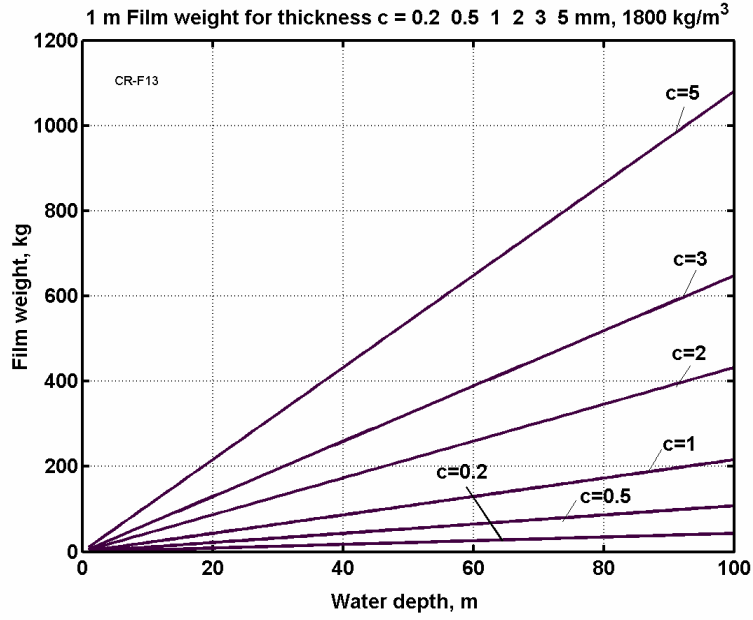

**Fig. 8**. One meter Film weight via the deep of dam and film thickness $c$, density 1800 kg/m$^3$.

**6. The diameter** $d$ of the support cable is

$$T = \frac{Pl_2}{2}, \quad S = \frac{T}{\sigma}, \quad d = \sqrt{\frac{4S}{\pi}}, \qquad (6)$$

where $T$ is cable force, N; $l_2$ is distance between cable, m; $S$ is cross-section area, m$^2$.
Computation is presented in fig. 8. The total weight of support cable is

$$W_c \approx 2\gamma_c HSL/l_2, \quad W_a = \gamma_c SL, \qquad (7)$$

where $\gamma_c$ is cable density, kg/m$^3$; $L$ is length of dam, m; $W_a$ is additional (connection of seashore) cable, m. The cheapest currently marketed textile-suitable fiber has $\sigma = 620$ kg/mm$^2$ and specific gravity $\gamma = 1.8$ g/cm$^3$.

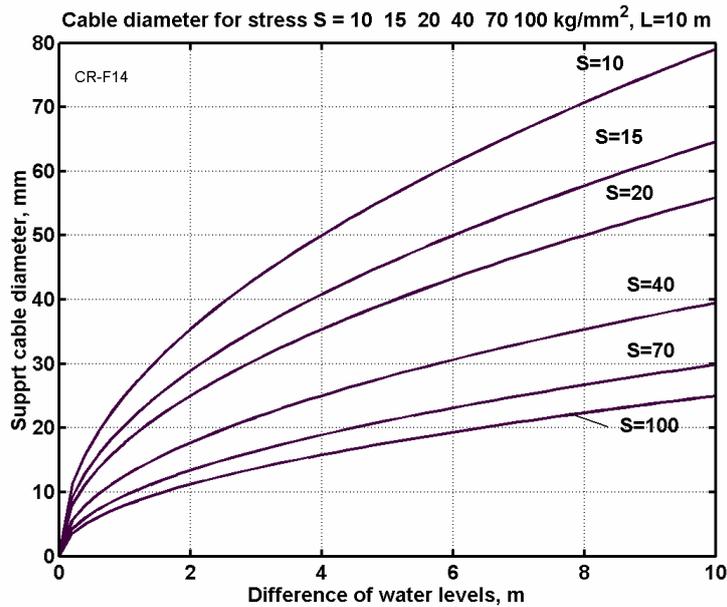

**Fig. 9**. Diameter of the support cable via water level differences and the safety tensile stress for every 10 m textile dam.

## 9. Conclusions

A drastic change in the State of California's climate—that is, a drying future period or a future wetter period compared to today's alleged "normal"—may necessitate the construction of a Golden Gate Textile Barrier. We have shown how this may be accomplished, but only in a preliminary Macro-engineering way.